\documentclass[twocolumn,floatfix,aps,eqsecnum,nofootinbib]{revtex4-2}
\usepackage{graphicx}
\usepackage{amsmath}
\usepackage{amssymb}
\usepackage{bm}
\usepackage{hyperref}

\begin{document}

\title{Lattice fermion simulation of spontaneous time-reversal symmetry breaking\\
in a helical Luttinger liquid}
\author{V. A. Zakharov}
\affiliation{Instituut-Lorentz, Universiteit Leiden, P.O. Box 9506, 2300 RA Leiden, The Netherlands}
\author{J. S\'{a}nchez Fern\'{a}n}
\affiliation{Instituut-Lorentz, Universiteit Leiden, P.O. Box 9506, 2300 RA Leiden, The Netherlands}
\author{C. W. J. Beenakker}
\affiliation{Instituut-Lorentz, Universiteit Leiden, P.O. Box 9506, 2300 RA Leiden, The Netherlands}

\date{January 2026}

\begin{abstract}
We extend a recently developed ``tangent fermion'' method to discretize the Hamiltonian of a helical Luttinger liquid on a one-dimensional lattice, including two-particle backscattering processes that may open a gap in the spectrum. The fermion-doubling obstruction of the sine dispersion is avoided by working with a tangent dispersion, preserving the time-reversal symmetry of the Hamiltonian. The numerical results from a tensor network calculation on a finite lattice confirm the expectation from infinite-system analytics, that a gapped phase with spontaneously broken time-reversal symmetry emerges when the Fermi level is tuned to the Dirac point and the Luttinger parameter crosses a critical value.
\end{abstract}
\maketitle

\section{Introduction}
\label{intro}

The edge of a quantum spin Hall insulator supports a time-reversally symmetric, one-dimensional metallic state in which spin and momentum are locked \cite{Has10,Mac11}. This gapless phase realizes a helical Luttinger liquid \cite{Wu06,Xu06,Lez12,Kai14,Lev09,Neu11,Li15,Dol16,Stu20,Hsu21,Jia22}, with time-reversal operation ${\cal T}$ squaring to $-1$, which is a distinct symmetry class from chiral or spinless Luttinger liquids \cite{Gia03}. A $\mathbb{Z}_2$ topological invariant forbids single-particle backscattering and prevents gap opening for the massless low-energy excitations \cite{Kan05}. The topological protection is lost if the edge is discretized with local couplings on a strictly one-dimensional (1D) lattice. In this sense the helical Luttinger liquid is ``holographic'' \cite{Wu06}: The 1D continuum theory emerges on the boundary of a 2D lattice.

Recent work \cite{Zak24a,Zak24b,Hae24} has shown that a 1D lattice formulation of the helical Luttinger liquid becomes possible for a \textit{nonlocal} discretization of the momentum operator \cite{Sta82}, resulting in a tangent dispersion $E\propto\tan(ka/2)$ instead of the usual sine dispersion $E\propto\sin ka$ (for lattice constant $a$ and momentum $|k|<\pi/a$ in the first Brillouin zone). Such \textit{tangent fermions} preserve time-reversal symmetry and spin-momentum locking, remaining in the symmetry class of a helical liquid without invoking a 2D bulk \cite{Bee23}. Most importantly, the nonlocality of the discretized Hamiltonian $H\propto \tan(ka/2)$ can be removed by working with a generalized eigenproblem \cite{Pac21}, of the form $P\psi=EQ\psi$ with local operators $P\propto\sin ka$ and $Q\propto 1+\cos ka$.

In the present work we extend the calculations of Ref.\ \onlinecite{Zak24b} to include more general two-particle interactions --- both gap-preserving and gap-opening. It is known from bosonization analyses \cite{Wu06,Xu06}, that strong interactions near half-filling (when the Fermi level is close to the Dirac point) can produce a pair of degenerate ground states, exchanged under $\cal T$, which gap the excitation spectrum. Here we recover that effect of spontaneous time-reversal symmetry breaking in a 1D lattice simulation, as an alternative to approaches that require a 2D lattice to realize the holographic liquid \cite{Hoh11,Hoh12,Ma24,Son24,Son25}. 

\section{Tangent fermion Luttinger liquid}

\subsection{Continuum formulation}

We summarize the continuum Hamiltonian of a helical Luttinger liquid \cite{Wu06,Xu06,Lez12,Kai14}.

We consider the edge of a quantum spin Hall insulator, along the $x$-axis, with fermionic field operators $\psi_\uparrow(x),\psi_\downarrow(x)$. Spin-up and spin-down electrons are counter-propagating (helical motion, with spin-momentum locking). We assume that spin-up electrons move to the right and spin-down electrons move to the left. The free Hamiltonian in the continuum is
\begin{equation}
H_0=v_{\rm F}\int dx\,\bigl(\psi^\dagger_\uparrow\,{p}\,\psi_\uparrow^{\vphantom{\dagger}}-\psi^\dagger_\downarrow\,{p}\, \psi_\downarrow^{\vphantom{\dagger}}\bigr),\label{H0def}
\end{equation} 
with $v_{\rm F}$ the Fermi velocity and ${p}=-i\hbar\partial/\partial x\equiv-i\hbar\partial_x$ the momentum operator.

The time-reversal operation ${\cal T}$, squaring to $-1$, transforms the field operators as
\begin{equation}
\psi_\uparrow(x)\mapsto\psi_\downarrow(x),\;\;\psi_\downarrow(x)\mapsto -\psi_\uparrow(x).
\end{equation}
Because also $p\mapsto-p$, the time-reversal operation leaves $H_0$ invariant. A single-particle spin-flip term, of the form
\begin{equation}
H_{\rm flip}=g_{\rm flip}\int dx\,\bigl[\psi^\dagger_\uparrow(x)\psi_\downarrow^{\vphantom{\dagger}}(x)+\psi^\dagger_\downarrow(x) \psi_\uparrow^{\vphantom{\dagger}}(x)\bigr],
\end{equation}
changes sign under ${\cal T}$ --- so it is forbidden if time-reversal symmetry is preserved (zero magnetic field, no magnetic impurities). Since the spin-flip changes the direction of motion, time-reversal symmetry prevents single-particle backscattering in the helical edge state.

Electron-electron interactions enable forward scattering processes, either within a single spin band (intra-band) or between the spin bands (inter-band). The corresponding interaction Hamiltonians are 
\begin{align}
&H_{\rm inter}=g_{\rm inter}\int dx\,\rho_\uparrow(x)\rho_\downarrow(x),\\
&H_{\rm intra}=g_{\rm intra}\int dx\,\bigl[\rho_\uparrow(x)\rho_\uparrow(x)+\rho_\downarrow(x)\rho_\downarrow(x)\bigr],\label{Hintradef}
\end{align}
with $\rho_\sigma(x)=\,:\!\psi^\dagger_\sigma(x)\psi_\sigma^{\vphantom{\sigma}}(x)\!:$ the normally ordered spin-$\sigma$ number density.\footnote{The product of field operators at the same position is regularized by a point-splitting operation. We will instead use a lattice regularization, where the operators are evaluated at neighbouring sites, see App.\  \ref{app_pointsplitting}.}

Interactions also enable backscattering without breaking time-reversal symmetry, via two types of ``Umklapp'' processes, in which one particle (U1) or two particles (U2) change direction of motion:
\begin{align}
H_{\rm U2}={}& g_{\rm U2}\int dx\,[\psi^\dagger_\uparrow(x)\psi_\downarrow^{\vphantom{\dagger}}(x)\psi^\dagger_\uparrow(x)\psi_\downarrow^{\vphantom{\dagger}}(x)\nonumber\\
{}&+\psi^\dagger_\downarrow(x)\psi_\uparrow^{\vphantom{\dagger}}(x)\psi^\dagger_\downarrow(x)\psi_\uparrow^{\vphantom{\dagger}}(x)],\label{HU2def}\\
H_{\rm U1}={}&ig_{\rm U1}\int dx\, [\rho_\uparrow(x)-\rho_\downarrow(x)]\nonumber\\
&\times[\psi^\dagger_\downarrow(x)\psi^{\vphantom{\dagger}}_\uparrow(x)-\psi^\dagger_\uparrow(x)\psi^{\vphantom{\dagger}}_\downarrow(x)] .\label{HU1def}
\end{align}

\subsection{Lattice formulation}

On a 1D lattice we discretize the momentum operator $p=-i\hbar \partial/\partial x$ by means of Stacey's long-range finite difference \cite{Sta82},
\begin{equation}
p\psi(x)\mapsto -\frac{2i\hbar}{a}\sum_{n=1}^\infty(-1)^n[\psi(x-na)-\psi(x+na)]. \label{Staceyderivative}
\end{equation}
The non-locality can be removed by noting that Eq.\ \eqref{Staceyderivative} can be written equivalently as \cite{Pac21}
\begin{equation}
p\psi\mapsto \frac{2\hbar}{a}\frac{\sin\hat{k}a}{1+\cos\hat{k}a}\psi,\;\;e^{i\hat{k}a}f(x)=f(x+a).
\end{equation}
The nonlocal Schr\"{o}dinger equation $H_0\psi=E\psi$ therefore transforms into a \textit{local} generalized eigenproblem upon substitution of $\psi=(1+\cos \hat{k}a)\psi'$. 

These lattice fermions (``tangent fermions'') have a tangent dispersion relation
\begin{equation}
E(k)=(2\hbar v_{\rm F}/a)\tan(ka/2),
\end{equation}
with a single Dirac point in the Brillouin zone $|k|<\pi/a$ (no fermion doubling). On a lattice of $L$ sites, we impose periodic boundary conditions with odd $L$, or anti-periodic boundary conditions with even $L$, so that the discrete wave numbers avoid the pole at $|k|=\pi/a$.

The free Hamiltonian \eqref{H0def} with discretization \eqref{Staceyderivative} becomes
\begin{equation}
\begin{split}
&H_0=\sum_{n>m} t_{nm}\bigl(c_{n\uparrow}^\dagger c_{m\uparrow}^{\vphantom{\dagger}}-c_{n\downarrow}^\dagger c_{m\downarrow}^{\vphantom{\dagger}}\bigr)+\text{H.c.},\\
&t_{nm}=2it_0 (-1)^{n-m},\;\;t_0=\hbar v_{\rm F}/a,
\end{split}
\label{H0lattice}
\end{equation}
in terms of the fermion creation and annihilation operators $c^{\vphantom{\dagger}}_{n\sigma},c^\dagger_{n,\sigma}$.
The inter-band and intra-band interactions are
\begin{align}
&H_{\rm inter}=t_{\rm inter}\sum_{n} \rho_{n\uparrow}\rho_{n\downarrow},\;\;\rho_{n\sigma}=c_{n\sigma}^\dagger c_{n\sigma}^{\vphantom{\dagger}},\\
&H_{\rm intra}=t_{\rm intra}\sum_{n}\bigl( \rho_{n\uparrow}\rho_{n+1,\uparrow}+\rho_{n\downarrow}\rho_{n+1,\downarrow}\bigr),\label{Hintralattice}\\
&t_{\rm inter}=g_{\rm inter}/a,\;\;t_{\rm intra}=\tfrac{1}{2}g_{\rm intra}/a.\label{tginterintra}
\end{align}
The factor of two in the relation between $t_{\rm intra}$ and $g_{\rm intra}$ appears when the point-splitting operation is regularized on the lattice (see App.\ \ref{app_pointsplitting}).

Similarly, the lattice regularization of the backscattering processes \eqref{HU2def} and \eqref{HU1def} is
\begin{align}
H_{\rm U2}={}&t_{\rm U2}\sum_{n}\bigl[c_{n\uparrow}^\dagger c_{n\downarrow}^{\vphantom{\dagger}}c_{n+1,\uparrow}^\dagger c_{n+1,\downarrow}^{\vphantom{\dagger}}\nonumber\\
&{}+c_{n\downarrow}^\dagger c_{n\uparrow}^{\vphantom{\dagger}}c_{n+1,\downarrow}^\dagger c_{n+1,\uparrow}^{\vphantom{\dagger}}\bigr],\label{HU2lattice}\\
H_{\rm U1}={}&it_{\rm U1}\sum_n\bigl[ (\rho_{n+1,\uparrow}-\rho_{n+1,\downarrow})(c^\dagger_{n\uparrow}c_{n\downarrow}^{\vphantom{\dagger}}-c^\dagger_{n\downarrow}c_{n\uparrow}^{\vphantom{\dagger}})\nonumber\\
{}+(&\rho_{n\downarrow}-\rho_{n\uparrow})(c^\dagger_{n+1,\uparrow}c_{n+1,\downarrow}^{\vphantom{\dagger}}-c^\dagger_{n+1,\downarrow}c_{n+1,\uparrow}^{\vphantom{\dagger}})\bigr],\label{HU1lattice}\\
t_{\rm U2}={}&\tfrac{1}{2}g_{\rm U2}/a,\;\;t_{\rm U1}=\tfrac{1}{2}g_{\rm U1}/a.
\end{align}
To check the time-reversal invariance, note that $\cal T$ transforms $\rho_{n\uparrow}\mapsto\rho_{n\downarrow}$ and $ic^\dagger_{n\downarrow}c_{n\uparrow}^{\vphantom{\dagger}}\mapsto ic^\dagger_{n\uparrow}c_{n\downarrow}^{\vphantom{\dagger}}$.

\subsection{Matrix product operator representation}

The lattice formulation makes it possible to treat the interactions numerically on a tensor network \cite{Ver04,Zwo04,Sch11}, in which the Hamiltonian is represented by a matrix product operator (MPO): A product of matrices $M^{(n)}$ of fermion operators acting only on site $n$. Such an approach is efficient if the rank of each matrix (the bond dimension) is fixed --- not growing with the number of sites $L$. 

An exact MPO representation at fixed bond dimension is possible even if the Hamiltonian is nonlocal, provided that the generalized eigenproblem is local \cite{Zak24b,Hae24}. Thus the tangent fermion Hamiltonian \eqref{H0lattice} has MPO representation \cite{Zak24b}
\begin{subequations}
\label{MPOtangenthelical}
\begin{align}
&H_{0}=2it_0[M^{(1)}M^{(2)}\cdots M^{(L)}]_{1,6},\\
&M^{(n)}=\begin{pmatrix}
1&c_{n\uparrow}&c_{n\uparrow}^\dagger&c_{n\downarrow}&c_{n\downarrow}^\dagger&0\\
0&-1&0&0&0&c_{n\uparrow}^\dagger\\
0&0&-1&0&0&c_{n\uparrow}\\
0&0&0&-1&0&-c_{n\downarrow}^\dagger\\
0&0&0&0&-1&-c_{n\downarrow}\\
0&0&0&0&0&1
\end{pmatrix},
\end{align}
\end{subequations}
of $L$-independent bond dimension 6.

The tensor network calculation then proceeds as described in Ref.\ \onlinecite{Zak24b}. Fermionic statistics is implemented by means of the Jordan-Wigner transformation. The ground state wave function $\Psi$ is represented by a matrix product state (MPS) and the DMRG algorithm (density matrix renormalization group \cite{Sch11}) from the TeNPy Library \cite{tenpy} is used to variationally minimize $\langle\Psi|H|\Psi\rangle/\langle\Psi|\Psi\rangle$. The bond dimension $\chi$ of the MPS is increased until convergence is reached. Our computer codes are available in a repository \cite{Zenodo}.

The Luttinger liquid is simulated at zero temperature and at fixed particle number $N=N_\uparrow+N_\downarrow$ (canonical ensemble). The half-filled band (zero chemical potential, Fermi level aligned with Dirac point) corresponds to $N=L$. Anti-periodic boundary conditions are implemented with $L$ even, so that $N_\uparrow=N_\downarrow=L/2$ are equal at half-filling.\footnote{In Ref.\ \onlinecite{Zak24b} we took periodic boundary conditions with $L$ odd. In that case a half-filled band corresponds to different $N_\uparrow=(L+1)/2$ and $N_\downarrow=(L-1)/2$.} 

Spurious oscillations of the wave function on even and odd lattice sites are removed by averaging the fermionic operators $c_{n\sigma}$ over nearby lattice sites,
\begin{equation}
\bar{c}_{n\sigma}=\tfrac{1}{2}c_{n\sigma}+\tfrac{1}{2}c_{n+1,\sigma}.
\end{equation}
Smoothed correlators are then obtained from the ground state expectation values
\begin{equation}
\begin{split}
&\bar{C}_\sigma(n,m)=\langle\bar{c}_{n\sigma}^\dagger\bar{c}_{m\sigma}\rangle,\\
&\bar{R}_\alpha(n,m)=\tfrac{1}{4}\langle \bm{\bar{c}}^\dagger_n{\sigma}_\alpha \bm{\bar{c}}_n \;\;\bm{\bar{c}}^\dagger_m{\sigma}_\alpha \bm{\bar{c}}_m\rangle,
\end{split}
\label{correlators}
\end{equation}
defined in terms of the spinor $\bm{c}=(c_\uparrow,c_\downarrow)$ and Pauli matrices ${\sigma}_\alpha$.

\section{Results for forward scattering}

If we only include forward scattering ($t_{\rm U1}=0=t_{\rm U2}$), the bosonization theory of an infinite Luttinger liquid \cite{Gia03,vanDelft98} gives a power law decay of the zero-temperature, zero-chemical-potential correlators,\footnote{Notice the factor of four difference in the definitions of $\kappa_{\rm inter}$ and $\kappa_{\rm intra}$: one factor of two comes from the sum over both spin directions in Eq.\ \eqref{Hintralattice}, the other factor of two comes from Eq.\ \eqref{tginterintra}.}
\begin{equation}
\begin{split}
&C_\sigma(x,x')\propto |x-x'|^{-(1/2)(K+1/K)},\\
&R_x(x,x'),\;\;R_y(x,x')\propto |x-x'|^{-2K},\\
&R_z(x,x')\propto |x-x'|^{-2},
\end{split}
\label{powerlaw}
\end{equation}
with Luttinger parameter $K$ given by
\begin{equation}
\begin{split}
&K=\sqrt{\frac{1+\kappa_{\rm intra}-\kappa_{\rm inter}}{1 +\kappa_{\rm intra}+\kappa_{\rm inter}}},\\
&\kappa_{\rm inter}=\frac{t_{\rm inter}}{2\pi t_0},\;\;\kappa_{\rm intra}=\frac{2t_{\rm intra}}{\pi t_0}.
\end{split}
\label{Kdef}
\end{equation}
The Fermi velocity is renormalized according to
\begin{equation}
v=v_{\rm F}\sqrt{(1+\kappa_{\rm intra})^2-\kappa_{\rm inter}^2}.\label{vrenormalized}
\end{equation}

We consider repulsive inter-band interactions, $t_{\rm inter}>0$, when $K\in(0,1)$. The transverse spin correlators $R_x$ and $R_y$ then decay more slowly than the $1/x^2$ decay expected from a Fermi liquid. If there are only intra-band interactions ($t_{\rm inter}=0$) one has $K=1$ and the correlators decay as for free electrons.

\begin{figure}[tb]
\centerline{\includegraphics[width=0.8\linewidth]{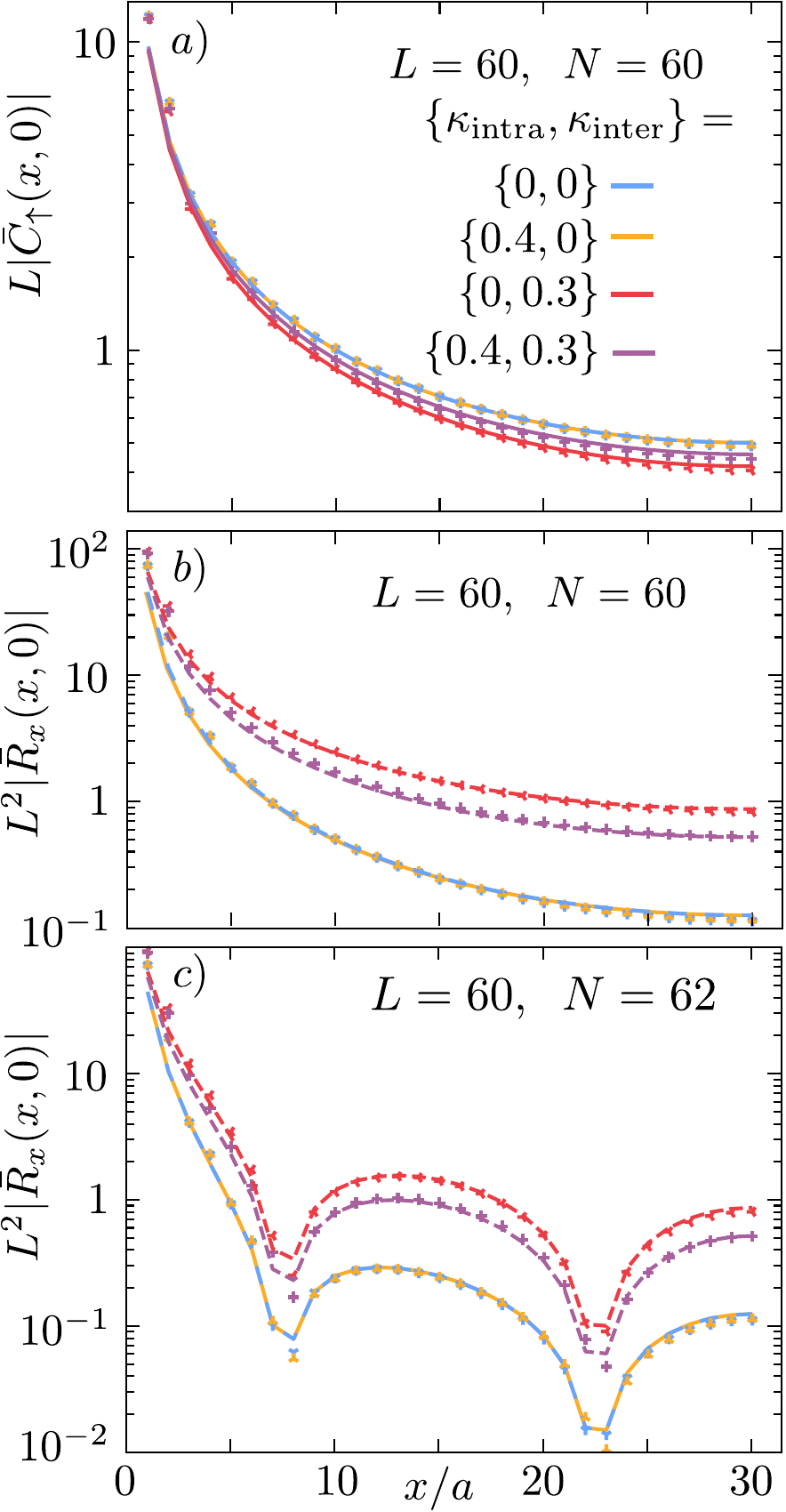}}
\caption{Absolute value of the propagator $\bar{C}_\uparrow$ and transverse spin correlator $\bar{R}_x$, defined in Eq. \eqref{correlators}, calculated in the tensor network of $L= 60$ sites (MPS bond dimension $\chi= 8192$). The data points are computed from the tangent discretization of the Luttinger Hamiltonian, for free fermions and for a repulsive intra-band and inter-band interaction. Only forward scattering is included in this figure ($t_{\rm U1}=0=t_{\rm U2})$. The curves are the analytical results in the continuum from bosonization theory. Panels a) and b) are for a half-filled band ($N=L)$. Panel c) shows  $\bar{R}_x$ away from half filling ($N=L+2$). The corresponding data for $\bar{C}_\uparrow$ is indistinguishable from panel a), so it is not included in the figure. For $\kappa_{\rm inter}=0$ the data is independent of $\kappa_{\rm intra}$, hence blue and orange data overlaps.
}
\label{fig_plotRC}
\end{figure}

Numerical results are shown in Fig.\ \ref{fig_plotRC} (data points). The curves are the continuum bosonization formulas (including finite-size corrections, see App.\ \ref{app_bosonization}). These results generalize those presented in Ref.\ \onlinecite{Zak24b} by the inclusion of intra-band scattering. We show both the case of a half-filled band and results away from half-filling. The lattice numerics is in very good agreement with the continuum analytics, without any fit parameter.

\section{Effects of backscattering}

\subsection{Gap opening}

\begin{figure}[tb]
\centerline{\includegraphics[width=1\linewidth]{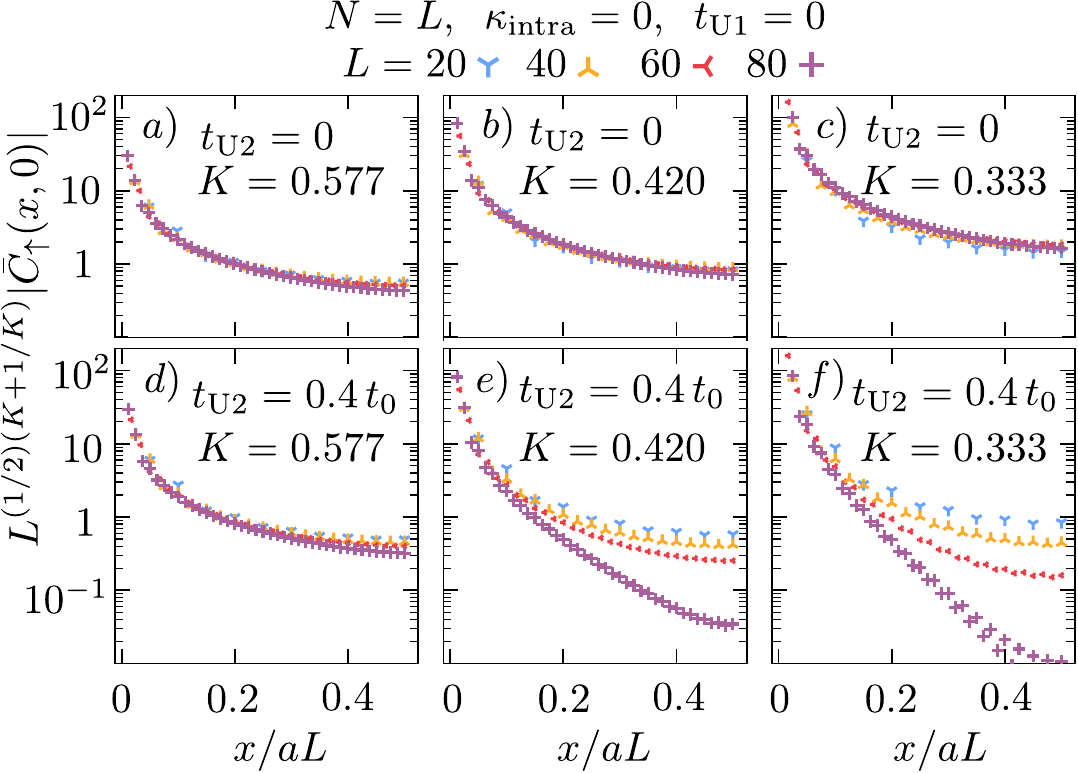}}
\caption{Absolute value of the propagator $\bar{C}_\uparrow$ calculated in the tangent fermion tensor network for different $L$ (MPS bond dimension $\chi= 2048$ for $L=20$, $\chi=4096$ for $L=40,60,80$). For all panels, the band is half-filled ($N=L$). The top row includes only inter-band forward scattering, for three different values of $\kappa_{\rm inter}=0.5,0.7,0.8$, corresponding to the values of the Luttinger parameter $K$ indicated in each panel. The lower row also includes two-particle backscattering ($t_{\rm U2}=0.4\,t_0$). The propagator is scaled by the power law \eqref{Cscaling}, so that in a gapless liquid curves for different $L$ collapse onto a single curve. This expected power law scaling breaks down in panels e) and f), with a crossover to an exponential decay indicating the opening of an excitation gap by two-particle backscattering for sufficiently strong forward scattering ($K<0.5$).
}
\label{fig_gap}
\end{figure}

\begin{figure}[tb]
\centerline{\includegraphics[width=0.8\linewidth]{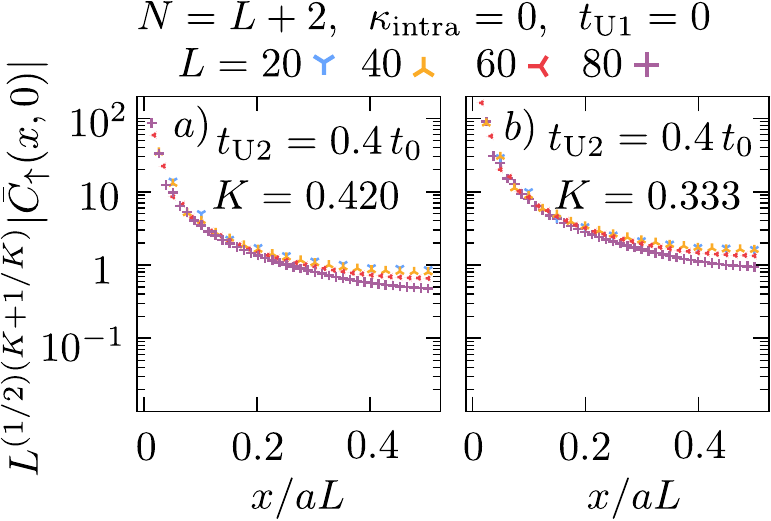}}
\caption{Same as Figs.\ \ref{fig_gap}e) and \ref{fig_gap}f), but away from half-filling ($N=L+2$). There is now no indication of a gap opening.
}
\label{fig_gapnonzeromu}
\end{figure}

\begin{figure}[tb]
\centerline{\includegraphics[width=1\linewidth]{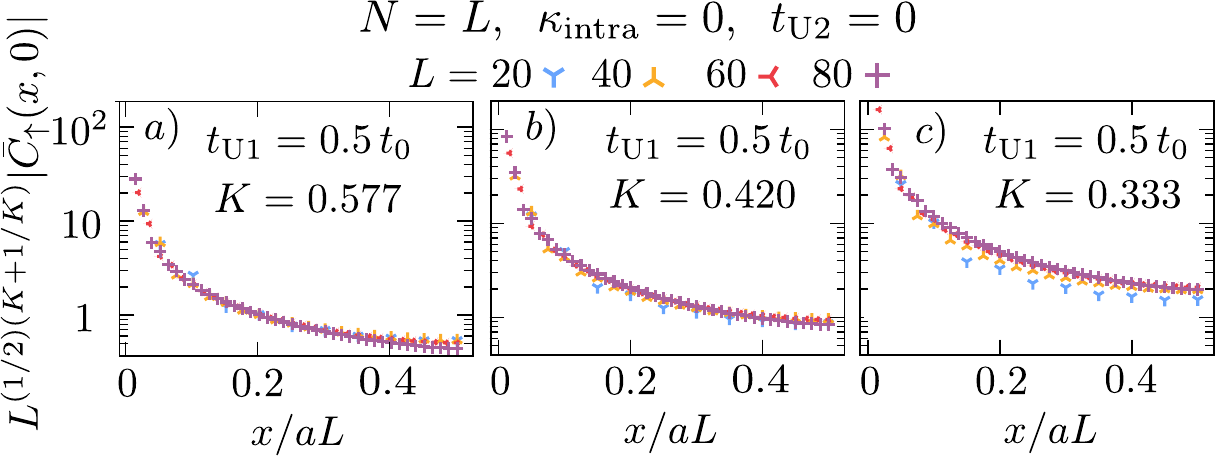}}
\caption{Same as Figs.\ \ref{fig_gap}d), \ref{fig_gap}e), and \ref{fig_gap}f), but for $t_{\rm U2}=0$ and nonzero $t_{\rm U1}$, to show that one-particle backscattering alone does not open a gap.
}
\label{fig_tU1}
\end{figure}

We next include backscattering, via nonzero $t_{\rm U1}$ or $t_{\rm U2}$. The renormalization group analysis \cite{Wu06,Kai14} tells us that a gap will open in a half-filled band in the $L\rightarrow\infty$ limit, provided that $t_{\rm U2}\neq 0$ and $K< K_c=1/2$. The numerical results for a finite system shown in Figs.\ \ref{fig_gap}, \ref{fig_gapnonzeromu}, \ref{fig_tU1} are in agreement with this expectation. 

In Fig.\ \ref{fig_gap} we see that the propagator for a half-filled band follows the power law scaling law\footnote{Since the decay of  $C_\sigma(x,0)$ with $x$ for $L\rightarrow\infty$ should be independent of $L$, the scaling law \eqref{Cscaling} implies the power law decay \eqref{powerlaw} with $x$. The function $f_C$ is given in App.\ \ref{app_bosonization}.}
\begin{equation}
C_\sigma(x,0)=L^{-(1/2)(K+1/K)}f_C(x/L)\label{Cscaling}
\end{equation}
of the gapless Luttinger liquid even in the presence of two-particle backscattering --- provided that the Luttinger parameter $K\gtrsim 1/2$. A break down of the power law scaling occurs for smaller $K$, when the decay of the propagator approaches the exponential decay of a gapped liquid. The range of $L$ that we can access numerically is not sufficiently large to precisely pinpoint the critical value of $K$, but the results are clearly consistent with $K_c=1/2$. 

Fig.\ \ref{fig_gapnonzeromu} shows that a half-filled band is essential for the gap opening and Fig.\ \ref{fig_tU1} shows that single-particle backscattering does not open a gap --- all consistent with the infinite-system analytics \cite{Wu06,Kai14}.

\subsection{Spontaneous time-reversal symmetry breaking}

The gap opening by interactions is expected to be accompanied by spontaneous time-reversal symmetry breaking \cite{Wu06}: The emergence of a pair of degenerate ground states exchanged under ${\cal T}$, each of which introduces a nonzero mass term 
\begin{equation}
M_\alpha(x)=\pm\tfrac{1}{2}\langle\psi^\dagger(x)\cdot\sigma_\alpha\cdot\psi(x)\rangle
\end{equation}
in the effective Hamiltonian. This term changes sign upon ${\cal T}$, which is allowed by time-reversally symmetric interactions because it appears with opposite sign in the two ground states. Which mass term is produced by the two-particle interaction depends on its sign \cite{Wu06}: $M_y$ if $t_{\rm U2}>0$ and $M_x$ if $t_{\rm U2}<0$.

\begin{figure}[tb]
\centerline{\includegraphics[width=1\linewidth]{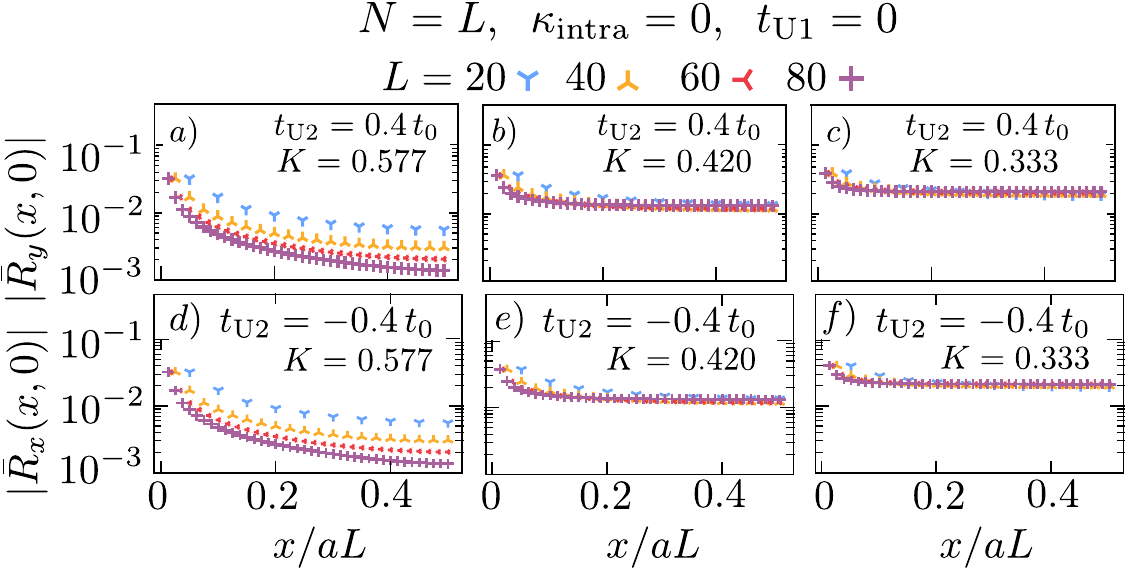}}
\caption{Top row: Same as Figs.\ \ref{fig_gap}d), \ref{fig_gap}e), and \ref{fig_gap}f), but for the transverse spin correlator $\bar{R}_y$ (no rescaling with $L$). The saturation of the correlator at a nonzero value in panels b), c) is a signature of spontaneous time-reversal symmetry breaking. The bottom row shows the same for negative $t_{\rm U2}$, when the saturated correlator is $ \bar{R}_x$ instead of $\bar{R}_y$.
}
\label{fig_mass}
\end{figure}

\begin{figure}[tb]
\centerline{\includegraphics[width=1\linewidth]{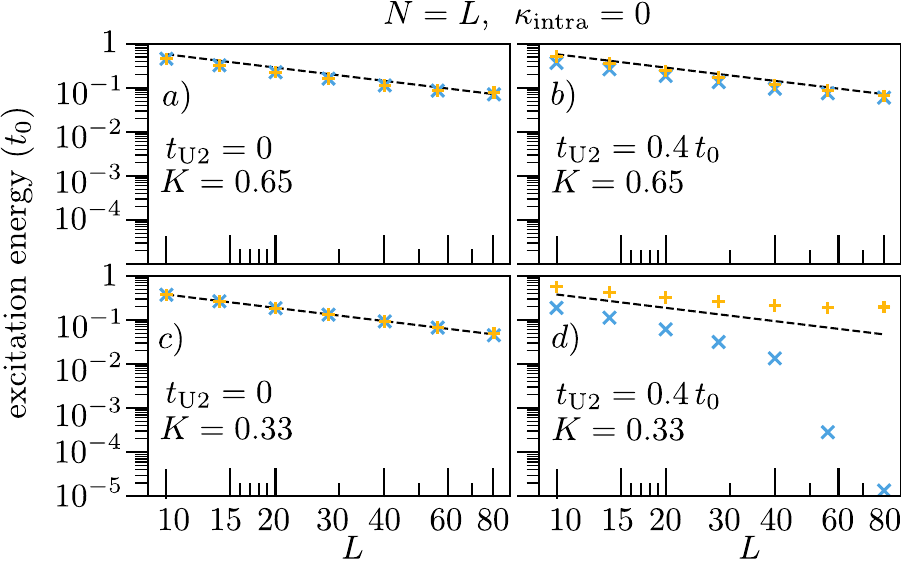}}
\caption{Log-log plot of the energy above the ground state of the first two excited states, as a function of system size $L$. Panels a,b,c show an excitation energy $\Delta E$ that decays as expected for a gapless system:  $\Delta E=2\pi \hbar v/L$ (dashed line), with velocity $v$ renormalized according to Eq.\ \eqref{vrenormalized}. Panel d) shows the signature of spontaneous time-reversal symmetry breaking due to a nonzero $t_{\rm U2}$ at $K<1/2$: One of the two excited states (blue data points) becomes degenerate with the ground state, while the other excited state (yellow data points) levels off at a nonzero $\Delta E$.
}
\label{fig_degeneracy}
\end{figure}

Since for large $x$ the transverse spin correlator factorizes,
\begin{equation}
R_\alpha(x,x')\rightarrow M_\alpha(x)M_\alpha(x'),\;\;|x-x'|\rightarrow\infty,
\end{equation}
we can test for the appearance of a nonzero $M_\alpha$ by checking whether the correlator $R_\alpha(x,0)$ saturates at a  nonzero value for large $x$. Fig.\ \ref{fig_mass} shows that this is indeed what happens, once $K$ drops below $1/2$. Finally, Fig.\ \ref{fig_degeneracy} shows the emergence of a ground-state degeneracy in the gapped regime.

\section{Conclusion}

It was shown in Refs.\ \onlinecite{Zak24a,Zak24b,Hae24} that a helical Luttinger liquid can be formulated and simulated directly on a one-dimensional lattice, without invoking the two-dimensional bulk that supports this ``holographic'' quantum liquid \cite{Wu06,Xu06}. By employing a tangent dispersion \cite{Sta82}, supporting a \textit{local} generalized eigenproblem \cite{Bee23,Pac21}, fermion doubling is avoided while time-reversal symmetry and spin-momentum locking are retained at the lattice level.

Only the gapless phase of the Luttinger liquid was explored in these studies \onlinecite{Zak24a,Zak24b,Hae24}, a phase where closed-form analytical results from bosonization theory are available \cite{Gia03}. Here we show that the same framework makes it possible to study numerically the predicted \cite{Wu06,Xu06,Lez12,Kai14} spontaneous time-reversal symmetry breaking and gap opening due to strong electron-electron interactions. Our results establish that the tangent-fermion regularization faithfully captures the low-energy physics of the helical Luttinger liquid, also in a regime where no closed-form analytical expressions are available.

\acknowledgments

We have benefited from discussions with A. S. Shankar.\\
This work was supported by the Netherlands Organisation for Scientific Research (NWO/OCW), as part of Quantum Limits (project number {\sc summit}.1.1016).

\appendix

\section{Lattice regularization of point splitting}
\label{app_pointsplitting}

\subsection{Intra-band scattering}

Point splitting \cite{Shankar,Wan19} is the operation that replaces the product of fermion fields at the same position by an infinitesimal displacement $\pm\epsilon$,
\begin{equation}
\psi_\sigma(x)\psi_\sigma(x)\mapsto\tfrac{1}{2}\psi_\sigma(x)\psi_\sigma(x+\epsilon)+\tfrac{1}{2}\psi_\sigma(x-\epsilon)\psi_\sigma(x).
\end{equation}
To first order in $\epsilon$, the right-hand-side inserts the derivative $\epsilon\psi(x)\partial_x\psi(x)$, which is how the point-splitting operation is usually introduced in the Luttinger Hamiltonian \cite{Lez12,Kai14}.

The product of density operators appearing in $H_{\rm intra}$ is split into
\begin{align}
\rho_\sigma(x)\rho_\sigma(x)\mapsto{}& \tfrac{1}{4}[\psi^\dagger_\sigma(x+\epsilon)\psi^\dagger_{\sigma}(x)+\psi^\dagger_{\sigma}(x)\psi^\dagger_\sigma(x-\epsilon)]\nonumber\\
&{}\times[\psi_\sigma(x)\psi_\sigma(x+\epsilon)+\psi_\sigma(x-\epsilon)\psi_\sigma(x)]\nonumber\\
={}&\tfrac{1}{4}\rho_\sigma(x)[\rho_\sigma(x+\epsilon)+\rho_\sigma(x-\epsilon)\nonumber\\
&\hspace*{-2cm}{}-\psi^\dagger_\sigma(x+\epsilon)\psi_\sigma(x-\epsilon)-\psi^\dagger_\sigma(x-\epsilon)\psi_\sigma(x+\epsilon)].
\end{align}

On the lattice we replace $\epsilon$ by the lattice spacing $a$. The integral over $x$ in Eq.\ \eqref{Hintradef} is replaced by the sum over sites of the operators $c_{n\sigma}=\sqrt{a}\psi_{\sigma}(x=na)$, resulting in
\begin{align}
a\int dx\,\rho_\sigma(x)\rho_\sigma(x)\mapsto\tfrac{1}{4}\sum_{n}\rho_{n\sigma}[\rho_{n+1,\sigma}+\rho_{n-1,\sigma}\nonumber\\
{}-c^\dagger_{n+1,\sigma}c_{n-1,\sigma}^{\vphantom{\dagger}}-c^\dagger_{n-1,\sigma}c_{n+1,\sigma}^{\vphantom{\dagger}}]\nonumber\\
=\tfrac{1}{2}\sum_{n}\rho_{n\sigma}(\rho_{n+1,\sigma}-1)+\tfrac{1}{4}\sum_{n}\rho_{n\sigma}J_{n\sigma},
\end{align}
where we have defined
\begin{equation}
J_{n\sigma}=2\rho_{n\sigma}-c^\dagger_{n+1,\sigma}c_{n-1,\sigma}^{\vphantom{\dagger}}-c^\dagger_{n-1,\sigma}c_{n+1,\sigma}^{\vphantom{\dagger}},
\end{equation}
and we have used that $\rho^2_{n\sigma}=\rho_{n\sigma}$.

The term $J_{n\sigma}$ is a discretized second derivative, in Fourier representation
\begin{equation}
\sum_n J_{n\sigma}=2\sum_k (1-\cos 2ka )c_{k\sigma}^\dagger c_{k\sigma}^{\vphantom{\dagger}},
\end{equation}
so it vanishes $\propto (ka)^2$ in the long-wave length limit. To make contact with the bosonisation theory we ignore this term in the main text, representing the intra-band scattering by
\begin{align}
H_{\rm intra}={}&\tfrac{1}{2}(g_{\rm intra}/a)\sum_{n,\sigma}\rho_{n\sigma}(\rho_{n+1,\sigma}-1)\nonumber\\
={}&t_{\rm intra}\sum_{n,\sigma}\rho_{n\sigma}\rho_{n+1,\sigma}+t_{\rm intra}N,
\end{align}
with $t_{\rm intra}=\tfrac{1}{2}g_{\rm intra}/a$ and $N=\sum_{n,\sigma}\rho_{n\sigma}$. Since we work in the canonical ensemble, at fixed particle number $N$, the offset $t_{\rm intra}N$ may be ignored, and we arrive at Eq.\ \eqref{Hintralattice} from the main text.

\subsection{Two-particle Umklapp scattering}

In a similar manner, the product of operators $\chi(x)=\psi^\dagger_{\uparrow}(x)\psi_\downarrow(x)$ appearing in $H_{\rm U2}$ is split into
\begin{align}
\chi(x)\chi(x)\mapsto{}&\tfrac{1}{4}[\psi^\dagger_\uparrow(x+\epsilon)\psi^\dagger_{\uparrow}(x)+\psi^\dagger_{\uparrow}(x)\psi^\dagger_\uparrow(x-\epsilon)]\nonumber\\
&{}\times[\psi_\downarrow(x)\psi_\downarrow(x+\epsilon)+\psi_\downarrow(x-\epsilon)\psi_\downarrow(x)]\nonumber\\
={}&\tfrac{1}{4}\chi(x)[\chi(x+\epsilon)+\chi(x-\epsilon)\nonumber\\
&\hspace*{-2cm}{}-\psi^\dagger_\uparrow(x+\epsilon)\psi_\downarrow(x-\epsilon)-\psi^\dagger_\uparrow(x-\epsilon)\psi_\downarrow(x+\epsilon)],
\end{align}
which on the lattice reduces to
\begin{align}
a\int dx\,\chi(x)\chi(x)\mapsto\tfrac{1}{4}\sum_{n}\chi_{n}[\chi_{n+1}+\chi_{n-1}\nonumber\\
{}-c^\dagger_{n+1,\uparrow}c_{n-1,\downarrow}^{\vphantom{\dagger}}-c^\dagger_{n-1,\uparrow}c_{n+1,\downarrow}^{\vphantom{\dagger}}]\nonumber\\
=\tfrac{1}{2}\sum_{n}\chi_{n}\chi_{n+1}+\tfrac{1}{4}\sum_{n}\chi_{n}K_{n},\\
K_{n}=2\chi_n-c^\dagger_{n+1,\uparrow}c_{n-1,\downarrow}^{\vphantom{\dagger}}-c^\dagger_{n-1,\uparrow}c_{n+1,\downarrow}^{\vphantom{\dagger}},
\end{align}
with $\chi_n=c_{n\uparrow}^\dagger c_{n\downarrow}^{\vphantom{\dagger}}$ and $\chi_n^2=0$.

The term $K_n$ is again a second derivative,
\begin{equation}
\sum_n K_{n}=2\sum_k (1-\cos 2ka )c_{k\uparrow}^\dagger c_{k\downarrow}^{\vphantom{\dagger}},
\end{equation}
which becomes irrelevant in the long-wave length regime. We thus represent the two-particle Umklapp term \eqref{HU2def} by the lattice regularization
\begin{equation}
H_{\rm U2}=\tfrac{1}{2}(g_{\rm U2}/a)\sum_{n}\bigl[c_{n\uparrow}^\dagger c_{n\downarrow}^{\vphantom{\dagger}}c_{n+1,\uparrow}^\dagger c_{n+1,\downarrow}^{\vphantom{\dagger}}+\text{H.c.}\bigr],
\end{equation}
which is Eq.\ \eqref{HU2lattice} in the main text.

\subsection{Single-particle Umklapp scattering}

For the $H_{\rm U1}$ interaction \eqref{HU1def} the point splitting is of the form
\begin{align}
&\psi_\uparrow^\dagger(x)\psi_\uparrow^\dagger(x)\psi_\uparrow^{\vphantom{\dagger}}(x)\psi_\downarrow^{\vphantom{\dagger}}(x)\mapsto{}\nonumber\\
&\tfrac{1}{2}[\psi^\dagger_\uparrow(x+\epsilon)\psi_\uparrow^\dagger(x)+\psi_\uparrow^\dagger(x)\psi_\uparrow^\dagger(x-\epsilon)]\psi_\uparrow^{\vphantom{\dagger}}(x)\psi_\downarrow^{\vphantom{\dagger}}(x).
\end{align}
The corresponding lattice regularization is
\begin{align}
H={}&\tfrac{1}{2}(g_{\rm U1}/a)\sum_n\bigl[ (\rho_{n+1,\uparrow}+\rho_{n\downarrow})(ic^\dagger_{n\uparrow}c_{n+1,\downarrow}^{\vphantom{\dagger}}+\text{H.c.})\nonumber\\
&- (\rho_{n+1,\downarrow}+\rho_{n\uparrow})(ic^\dagger_{n+1,\uparrow}c_{n\downarrow}^{\vphantom{\dagger}}+\text{H.c.})\bigr].\label{HU1app}
\end{align}

To implement this as a matrix product operator it is more convenient if the spin-flip operators act on the same site. For that purpose we make the replacements $c_{n\sigma}\leftrightarrow c_{n+1,\sigma}$, with an error that vanishes as $ka$ in the long-wave length limit. This gives
\begin{align}
H_{\rm U1}={}&\tfrac{1}{2}(g_{\rm U1}/a)\sum_n\bigl[ (\rho_{n\downarrow}-\rho_{n\uparrow})(ic^\dagger_{n+1,\uparrow}c_{n+1,\downarrow}^{\vphantom{\dagger}}+\text{H.c.})\nonumber\\
&+(\rho_{n+1,\uparrow}-\rho_{n+1,\downarrow})(ic^\dagger_{n\uparrow}c_{n\downarrow}^{\vphantom{\dagger}}+\text{H.c.})\bigr],
\end{align}
which is Eq.\ \eqref{HU1lattice} in the main text.

\section{Bosonization results with finite-size corrections}
\label{app_bosonization}

If there is only forward scattering ($g_{\rm U1}=0=g_{\rm U2}$) the correlators in the continuum can be calculated analytically from bosonization theory \cite{Gia03,vanDelft98}. To compare with the numerics on a finite lattice we need to include finite-size corrections. 

In the canonical ensemble at zero temperature the propagator $C_\sigma$ and transverse spin correlators $R_x=R_y$ are given by \cite{Zak24a,Zak24b}
\begin{subequations}
\begin{align}
&C_\sigma(x,0)=\frac{\sigma e^{  \delta N_{\sigma}(2\pi ix/L)}}{2\pi ia_{*}|(L/\pi a_{*})\sin(\pi x/L)|^{(1/2)(K+1/K)}},\\
	&R_x(x,0)
	= \frac{ \cos\bigl(2\pi(\delta N_\uparrow+\delta N_\downarrow)x/L\bigr)}{2(2\pi a_{*})^2|(L/\pi a_{*})\sin(\pi x/L)|^{2K}},
\end{align}
\end{subequations}
 in a system of length $L$ with antiperiodic boundary conditions and short-distance (UV) regularization constant $a^\ast$. The number $\delta N_\sigma$ is the number of spin-$\sigma$ electrons relative to a half-filled band. The Luttinger parameter $K$, dependent on the inter-band and intra-band forward scattering strengths, is given by Eq.\ \eqref{Kdef}. 

For the comparison with a lattice calculation the lengths $x$ and $L$ are measured in units of the lattice constant $a$. A comparison \cite{Zak24b} of lattice and UV regularization gives $a=2a^\ast$.

\end{document}